\documentclass[8.5pt,twoside,twocolumn]{article}
\usepackage[super,sort&compress,comma]{natbib}
\usepackage{mhchem}
\usepackage{times,mathptmx}
\usepackage{sectsty}
\usepackage{balance}

\usepackage{graphicx} 
\usepackage{lastpage}
\usepackage[format=plain,justification=raggedright,singlelinecheck=false,font=small,labelfont=bf,labelsep=space]{caption}
\usepackage{fancyhdr}
\begin{document}

\thispagestyle{plain}
\fancypagestyle{plain}{
\fancyhead[L]{\includegraphics[height=8pt]{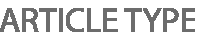}}
\fancyhead[C]{\hspace{-1cm}\includegraphics[height=20pt]{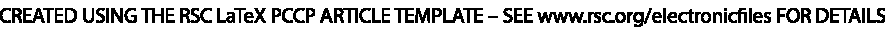}}
\fancyhead[R]{\includegraphics[height=10pt]{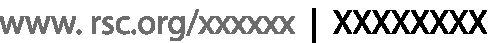}\vspace{-0.2cm}}
\renewcommand{\headrulewidth}{1pt}}
\renewcommand{\thefootnote}{\fnsymbol{footnote}}
\renewcommand\footnoterule{\vspace*{1pt}%
\hrule width 3.4in height 0.4pt \vspace*{5pt}}
\setcounter{secnumdepth}{5}

\makeatletter
\def\subsubsection{\@startsection{subsubsection}{3}{10pt}{-1.25ex plus -1ex minus -.1ex}{0ex plus 0ex}{\normalsize\bf}}
\def\paragraph{\@startsection{paragraph}{4}{10pt}{-1.25ex plus -1ex minus -.1ex}{0ex plus 0ex}{\normalsize\textit}}
\renewcommand\@biblabel[1]{#1}
\renewcommand\@makefntext[1]%
{\noindent\makebox[0pt][r]{\@thefnmark\,}#1}
\makeatother
\renewcommand{\figurename}{\small{Fig.}~}
\sectionfont{\large}
\subsectionfont{\normalsize}

\fancyfoot{}
\fancyfoot[LO,RE]{\vspace{-7pt}\includegraphics[height=9pt]{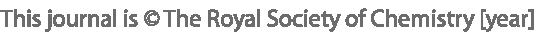}}
\fancyfoot[CO]{\vspace{-7.2pt}\hspace{12.2cm}\includegraphics{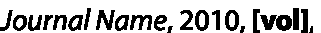}}
\fancyfoot[CE]{\vspace{-7.5pt}\hspace{-13.5cm}\includegraphics{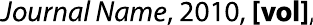}}
\fancyfoot[RO]{\footnotesize{\sffamily{1--\pageref{LastPage} ~\textbar  \hspace{2pt}\thepage}}}
\fancyfoot[LE]{\footnotesize{\sffamily{\thepage~\textbar\hspace{3.45cm} 1--\pageref{LastPage}}}}
\fancyhead{}
\renewcommand{\headrulewidth}{1pt}
\renewcommand{\footrulewidth}{1pt}
\setlength{\arrayrulewidth}{1pt}
\setlength{\columnsep}{6.5mm}
\setlength\bibsep{1pt}

\twocolumn[
  \begin{@twocolumnfalse}
\noindent\LARGE{\textbf{Local structure of semicrystalline P3HT films probed by nanofocused coherent x-rays$^\dag$}}
\vspace{0.6cm}

\noindent\large{\textbf{
Ruslan P.~Kurta,\textit{$^{a\ast}$}
Linda Grodd,\textit{$^{b}$}
Eduard Mikayelyan,\textit{$^{b}$}
Oleg Y.~Gorobtsov,\textit{$^{a,c}$}
Ivan A.~Zaluzhnyy,\textit{$^{a,d}$}
Ilaria Fratoddi,\textit{$^{e}$}
Iole Venditti,\textit{$^{f}$}
Maria Vittoria Russo,\textit{$^{f}$}
Michael Sprung,$^{a}$
Ivan A. Vartanyants$^{a,d\ddag}$
and
S.~Grigorian\textit{$^{b\text{\textsection\P}}$}}}\vspace{0.5cm}

\noindent\textit{\small{\textbf{Received Xth XXXXXXXXXX 20XX, Accepted Xth XXXXXXXXX 20XX\newline
First published on the web Xth XXXXXXXXXX 200X}}}

\noindent \textbf{\small{DOI: 10.1039/b000000x}}
\vspace{0.6cm}

\noindent \normalsize{
We present results of an x-ray study of structural properties of semicrystalline polymer films using nanofocused x-ray beam.
We applied the x-ray cross-correlation analysis (XCCA) to scattering data from blends of poly(3-hexylthiophene) (P3HT) embedded with gold nanoparticles (AuNPs).
Spatially resolved maps of orientational distribution of crystalline domains allow us to distinguish sample regions of
predominant face-on morphology, with a continuous transition to edge-on morphology.
The average size of crystalline domains was determined to be of the order of $10\;\textrm{nm}$.
As compared to pristine P3HT film, the P3HT/AuNPs blend is characterized by substantial ordering of crystalline domains,
which can be induced by Au nanoparticles.
The inhomogeneous structure of the polymer film is clearly visualized on the spatially resolved nanoscale 2D maps obtained using XCCA.
Our results suggest that the observed changes of the polymer matrix within crystalline regions
can be attributed to nanoconfinement in the presence of gold nanoparticles.
}
\vspace{0.5cm}
 \end{@twocolumnfalse}
  ]

\footnotetext{\dag~Electronic Supplementary Information (ESI) available. See DOI: 10.1039/b000000x/}


\footnotetext{\textit{$^{a}$~Deutsches Elektronen-Synchrotron DESY, Notkestra\ss e 85, D-22607 Hamburg, Germany. E-mail: ruslan.kurta@desy.de}}
\footnotetext{\textit{$^\ast$~Present address: European X-Ray Free-Electron Laser Facility, Albert-Einstein-Ring 19, D-22761 Hamburg, Germany.}}
\footnotetext{\textit{$^{b}$~Department of Physics, University of Siegen, Walter-Flex-Stra\ss e 3, D-57072 Siegen, Germany. }}
\footnotetext{\textit{$^{c}$~National Research Center ``Kurchatov Institute'', Kurchatov Square 1, 123182 Moscow, Russia.}}
\footnotetext{\textit{$^{d}$~National Research Nuclear University, ``MEPhI'', 115409 Moscow, Russia.}}
\footnotetext{\textit{$^{e}$~Department of Chemistry and Center for Nanotechnology for Engineering (CNIS), University of Rome Sapienza, P.le A. Moro 5, I-00185 Rome, Italy. }}
\footnotetext{\textit{$^{f}$~Department of Chemistry, University of Rome Sapienza, P.le A. Moro 5, I-00185 Rome, Italy. }}
\footnotetext{\textit{$^\ddag$~E-mail: ivan.vartaniants@desy.de }}
\footnotetext{\textit{$^\text{\textsection}$~E-mail: grigorian@physik.uni-siegen.de }}
\footnotetext{\textit{$^\text{\P}$~Experiments were designed by S.G. and I.A.V. and performed by S.G.,  L.G., E.M., R.K., O.Y.G. and M.S.; samples were prepared by L.G., E.M., S.G. in collaboration with I.F., I.V. and M.V.R.; data analysis and interpretation was performed by R.K., I.A.Z. and I.A.V.; the manuscript was written by R.K. and I.A.V.; all authors discussed the results and commented on the manuscript.}}


\section{Introduction}

Semicrystalline conjugated polymers are promising cost-effective candidates for organic electronic devices \cite{Kline2, Majewski, Coakley}.
Among different conjugated polymers, polythiophenes received increasing attention in recent years due
to their attractiveness for organic field-effect transistors (OFET) and solar cell applications \cite{Sirring, Salleo, Dang, Krebs, Kim1, Nagarjuna}.
Typical features of these polymers are mixtures of poor and well organized domains \cite{Kohn1, Brinkmann}.
Usually, well organized domains are addressed to be crystalline
and it is assumed that their presence strongly improves device performance \cite{Kline1, Zen, Ali, Tanigaki}.
Controlling the morphology and orientation of the crystalline domains in polymer films is a crucial step for the fabrication process,
since it determines key electronic properties of the material, such as charge carrier mobility and charge separation, and defines
the overall device performance \cite{Joshi, Tanigaki, Nagarjuna}.

\begin{figure}[h]
\centering
\includegraphics[width=8.3cm]{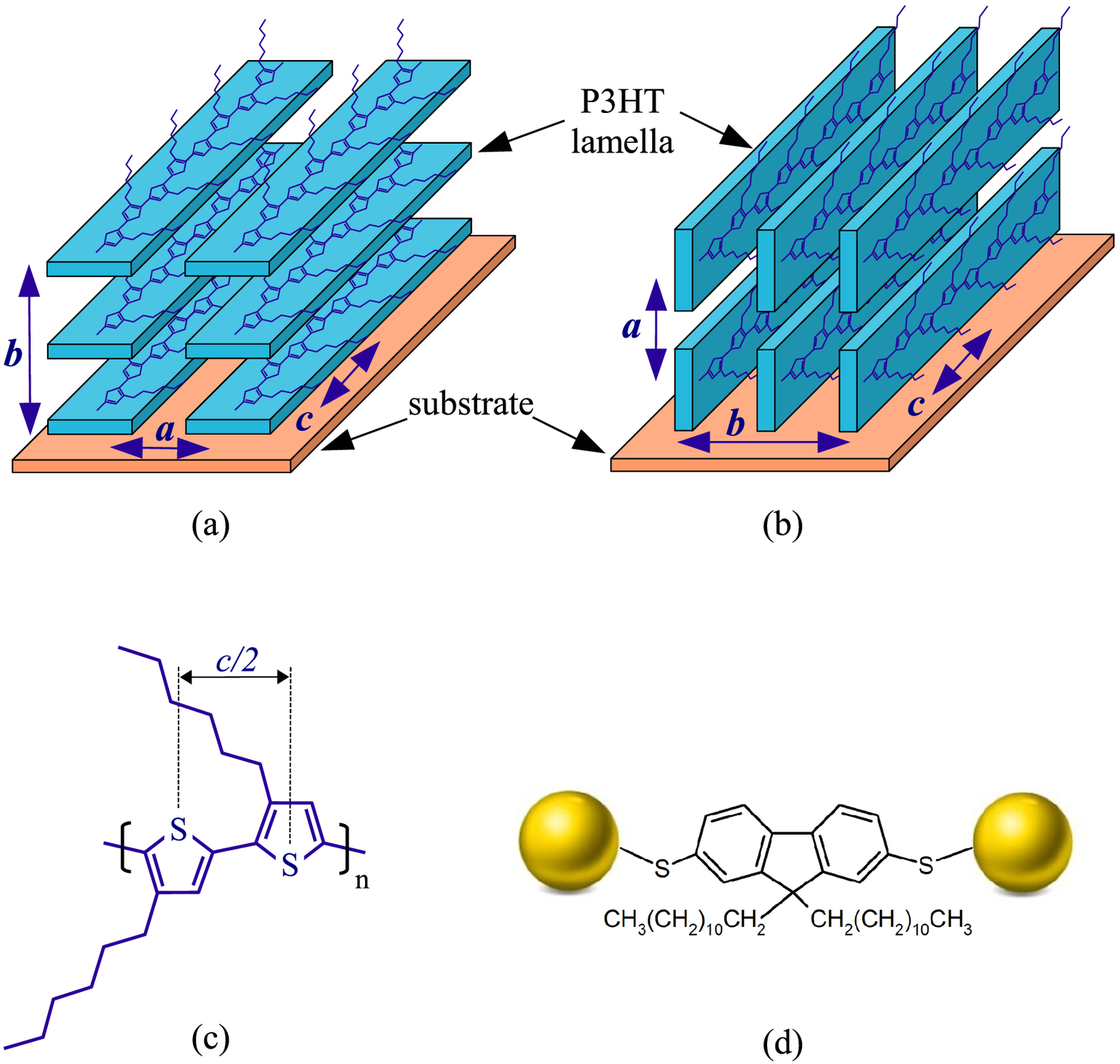}
\caption{Two types of predominant orientation of semicrystalline P3HT domains with respect to a substrate, the face-on (a) and edge-on (b) orientations.
(c) Repeating unit of the P3HT polymer chain. (d) Gold nanoparticles stabilized with fluorene derivatives (AuNPs-SFL).
\label{Fig:P3HT}}
\end{figure}

One of the most studied polythiophenes is poly(3-hexylthiophene) (P3HT) [see Fig.~\ref{Fig:P3HT}(c)]
with crystalline domain sizes varying from tens to few hundreds of nanometers,
depending on preparation techniques \cite{Prosa, Rahimi, Kohn1}.
In many cases two predominant morphologies, termed edge-on and face-on, can be observed for pristine P3HT \cite{Kim, Tanigaki}.
They are defined by a different orientation of crystalline domains with respect to the substrate.
P3HT lamellae stack parallel to the substrate in the case of face-on domains [Fig.~\ref{Fig:P3HT}(a)], and perpendicular in the case of edge-on domains [Fig.~\ref{Fig:P3HT}(b)].
The shortest distance between P3HT layers, called $\pi$-$\pi$ stacking distance, is equal to $b/2$, where $b$ is a unit cell parameter [Fig.~\ref{Fig:P3HT}(a),(b)].
Mixed orientation of domains also occurs, especially in non annealed samples.
Generally, it is assumed that the edge-on orientation of domains enhances OFET performance,
while the face-on orientation is favorable for photovoltaic applications \cite{Sirring, Salammal, Tanigaki}.

P3HT films can be prepared using various methods, for example, drop-casting, spin-coating, dip-coating or directional crystallization,
leading to a different degree of crystallinity and preferred orientation of P3HT domains \cite{Salleo, Shabi, Tanigaki, Brinkmann1}.
The formation of a particular morphology of the film is defined by macromolecular parameters (molecular weight, regioregularity, polydispersity,  etc),
sample growth conditions (e.g., temperature, coating speed) and post-processing (e.g., annealing)
\cite{Salleo, Kline1, Kline2, Hoppe, Salammal, Zhang, Shabi, Brinkmann1, Grigorian1, Zen, Tanigaki, Kohn, Kohn1, Kayunkid, Newbloom}.
It has been demonstrated that thermal annealing of the grown films results in better crystallinity and improved charge carrier mobility \cite{Zen, Ali, Salammal, Joshi1, Grigorian1}.

Together with pristine P3HT, its blends of various compositions emerge as another important class of hybrid materials with attractive structural and electronic properties \cite{Kohn, Moghaddam, Parashchuk}.
In this work we analyzed structural variations of the P3HT host matrix upon small addition of gold nanoparticles stabilized with fluorene derivatives (AuNPs-SFL) [see Fig.~\ref{Fig:P3HT}(d)].
Such systems have been extensively investigated \cite{Quintiliani, Battocchio1, Battocchio2} and are considered to be attractive for optoelectronic applications
due to their optical absorption and emission properties, as well as solubility \cite{Venditti1, Battocchio3, Ghosh}.


It is important to characterize the structure of the $\pi$-$\pi$ conjugated P3HT network on the nanoscale, since
charge transport and charge separation are governed by the nanoscale morphology of a polymer film \cite{Hoppe, Coakley, Clarke}.
Various techniques, such as x-ray, electron or neutron scattering, atomic force microscopy and transmission electron microscopy,
are used to experimentally characterize the sample morphology on different length scales \cite{Salleo, Kline1, Moghaddam, Salammal}.
To study structural variations in semicrystalline P3HT films on the nanoscale we performed spatially-resolved coherent x-ray scattering experiment with a nanofocused x-ray beam,
in combination with the x-ray cross-correlation analysis (XCCA)\cite{Wochner1, Altarelli, Kurta1, Kurta2, Kurta3}.
A nanosize x-ray probe provides access to the local structure of a polymer film, and XCCA gives information on orientational ordering on a larger length scales.
XCCA is a newly developed technique for structural characterization of partially ordered samples \cite{Kurta4, Grubel, Liu, Kurta5}.
Our approach provides information which is complementary to results of conventional small angle x-ray scattering (SAXS) or grazing incidence x-ray diffraction (GIXD) experiments \cite{Kohn, Grigorian1}.

This paper is organized as follows.
In the next section we provide a short theoretical basis for the XCCA technique applied for data analysis.
The description of the experimental setup, sample preparation and measurement scheme are presented in the third section of the paper.
In the fourth section results of the data analysis are presented, where the average and spatially resolved structural properties of the film are considered separately.
The conclusions section completes the paper.

\section{Theory}

\begin{figure*}[htb]
\centering
\includegraphics[width=14cm]{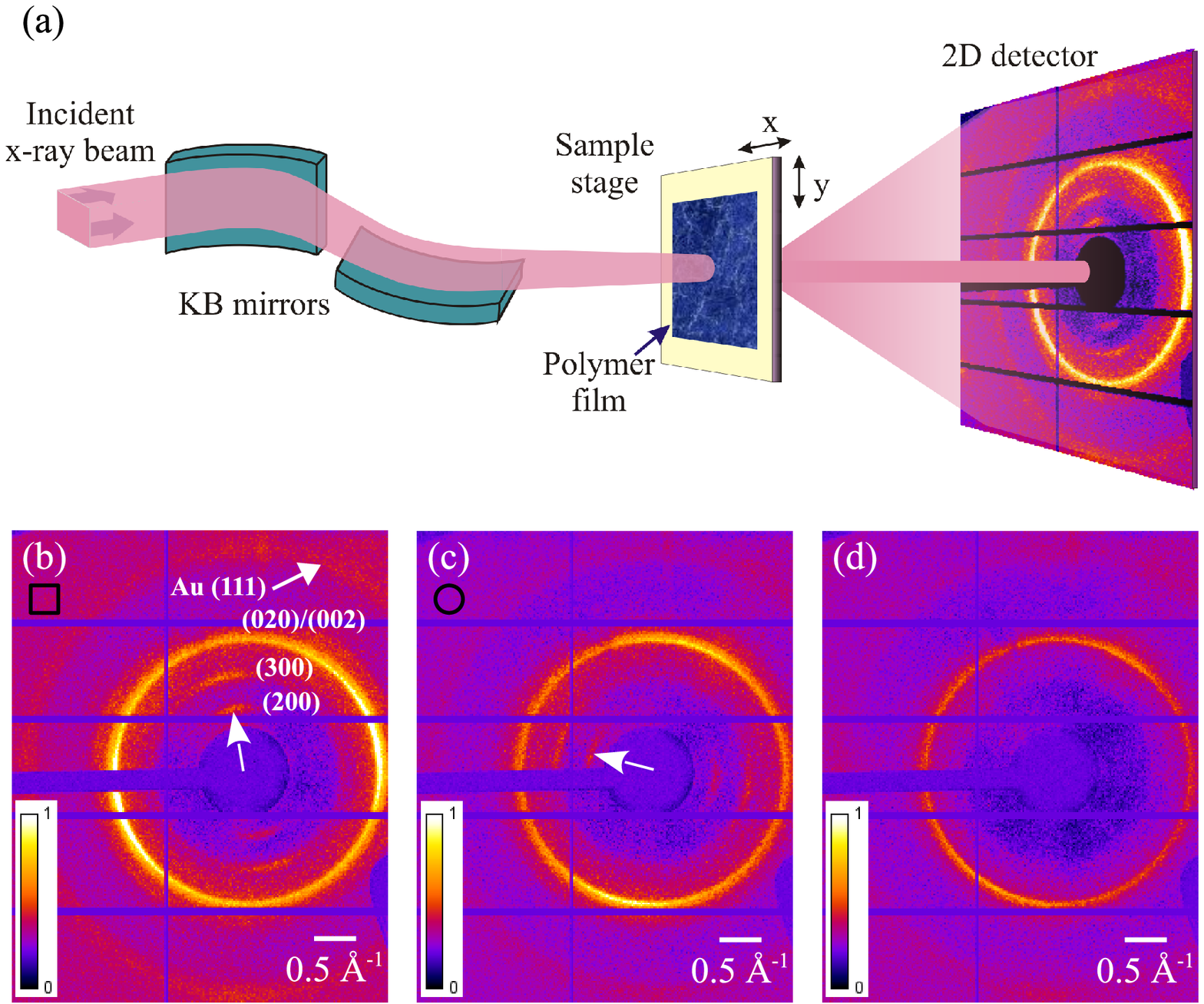}
\caption{(a) Geometry of the scattering experiment showing the focusing KB mirrors, sample, and 2D detector.
(b),(c),(d) Diffraction patterns measured at three different positions of the polymer film,
indicating the presence (b,c) and absence (d) of the face-on oriented P3HT domains. 
Diffraction patterns (b) and (c) were measured at the sample positions indicated in Fig.~\ref{Fig:VectorField}(c) by a black square and circle, correspondingly.
Arrows drawn from the center of the diffraction patterns (b) and (c) indicate orientation of the (200) peaks.
\label{Fig:ExpGeom}}
\end{figure*}

We consider coherent x-ray scattering from a partially ordered polymer film in transmission geometry, as it is shown in Fig.~\ref{Fig:ExpGeom}(a).
For each diffraction pattern the measured intensity distribution $I(\mathbf{ q})$ can be considered in a polar coordinate system,
where the momentum transfer vector $\mathbf{q}=(q,\varphi)$ is defined by the radial $q$ and angular $\varphi$ components.
The scattered intensity $I(q,\varphi)$ can be expanded into angular Fourier series as
\begin{equation}
I(q,\varphi)=I_{0}(q)+2\sum\limits_{n=1}^{\infty}|I_{n}(q)|\cos(n\varphi+\psi_{n}),\label{Eq:Cchi}
\end{equation}
where $I_{n}(q)$ are, in general, complex Fourier components of intensity $I(q,\varphi)$ with the amplitudes $|I_{n}(q)|$ and phases $\psi_{n}$.
In can be shown that $|I_{0}(q)|\equiv \left\langle I(q,\varphi)\right\rangle_{\varphi}$ for $n=0$,
where $\left\langle I(q,\varphi)\right\rangle_{\varphi}=1/(2\pi)\int_{0}^{2\pi}{I(q,\varphi)\textrm{d}\varphi}$
denotes the angular average of the scattered intensity around the ring of a radius $q$ \cite{Altarelli}.
The relation $I_{-n}(q)=I_{n}^{\ast}(q)$ was used in Eq.~(\ref{Eq:Cchi}) due to the fact that measured intensities are always real quantities.

XCCA enables direct determination of the Fourier components $I_{n}(q)$ from a set
of diffraction patterns measured at different positions on the sample \cite{Altarelli, Kurta1}.
The two-point cross-correlation function (CCF) $C^{i,j}(q,\Delta)$ can be defined as \cite{Kurta5}
\begin{equation}
C^{i,j}(q,\Delta) = \left\langle I^{i}(q,\varphi)I^{j}(q,\varphi+\Delta)\right\rangle_{\varphi},\label{Eq:CCF2}
\end{equation}
where $\Delta$ is the angular coordinate, and $I^{i}(q,\varphi)$ and $I^{j}(q,\varphi)$ indicate intensities measured on the $i$-th and $j$-th diffraction patterns, respectively.
We would like to note, that in Eq.~(\ref{Eq:CCF2}) the CCF $C^{i,j}(q,\Delta)$ can be determined on two different diffraction patterns $(i \neq j)$, or
individual diffraction pattern $(i=j)$.
In the latter case Eq.~(\ref{Eq:CCF2}) reduces to the well-known CCF $C^{i,i}(q,\Delta)$ often used in the literature \cite{Altarelli, Kurta1}.

The CCF $C^{i,j}(q,\Delta)$ can be analyzed using a Fourier series decomposition similar to Eq.~(\ref{Eq:Cchi})
\begin{equation}
C^{i,j}(q,\Delta)=C^{i,j}_{0}(q)+2\sum\limits_{n=1}^{\infty} |C^{i,j}_{n}(q)|\cos(n\Delta+\phi_{n}),\label{Eq:Cqn2}
\end{equation}
where $C^{i,j}_{n}(q)$ are the complex Fourier components of the CCF, with the corresponding amplitudes $|C^{i,j}_{n}(q)|$ and phases $\phi_{n}$.
It can be shown \cite{Altarelli, Kurta1,Kurta2,Kurta3,Kurta5} that $C^{i,j}_{n}(q)=I^{i\ast}_{n}(q)\cdot I^{j}_{n}(q)$, where
the Fourier components of intensity $I^{i}_{n}(q)$ and $I^{j}_{n}(q)$ are defined for the $i$-th and $j$-th diffraction patterns, respectively.
Particularly, in the case of $i=j$ the Fourier components $C^{i,i}_{n}(q)$ are real $(\phi_{n}=0)$, with
\begin{equation}
C^{i,i}_{n}(q)=|I^{i}_{n}(q)|^2,\label{Eq:Cqnii2}
\end{equation}
for $n\neq 0$, and $C^{i,i}_{0}(q)\equiv \langle I^{i}(q,\varphi)\rangle_{\varphi}^{2}$ for $n=0$.

It has been demonstrated \cite{Kurta1} that in the case of a system of identical particles with a substantial orientational order,
the Fourier components $C^{i,i}_{n}(q)$ calculated for different realizations of such a system qualitatively resemble a spectrum corresponding to a single particle.
This enables spatially resolved analysis of the diffraction data measured at each individual position of the sample.
In the case of a disordered system the Fourier components $C^{i,j}_{n}(q)$ fluctuate from position to position on the sample \cite{Altarelli,Kurta1,Kurta2,Kurta3}
and averaging over a large number $M$ of realizations of the system can be applied to get a statistical result,
\begin{equation}
\langle C^{i,j}_{n}(q) \rangle_{M}=1/M\sum_{i,j}^{M}C^{i,j}_{n}(q),
\label{Eq:CCFaver1}
\end{equation}
where the averaging of $C^{i,j}_{n}(q)$ is performed over $M$ pairs of diffraction patterns with $i \neq j$, or $M$ diffraction patterns in the case of $i=j$.
It can be shown that $\langle C^{i,j}_{n}(q)\rangle_{M}\to 0$ for $n\neq 0$ (see section 1.1 in the ESI$^\dag$), for a completely disordered system without
any angular correlation between diffraction patterns measured at different positions on the sample $(i\neq j)$.
Nonzero values of $\langle C^{i,j}_{n}(q)\rangle_{M}$ indicate the presence of background contributions in the scattered signal.

The experimentally determined spectrum $\langle C^{i,i}_{n}(q) \rangle_{M}$ may contain both, sample and background scattering contributions.
To remove background contribution from $\langle C^{i,i}_{n}(q) \rangle_{M}$ the following difference spectrum \cite{Kurta5} can be used for analysis,
\begin{equation}
\langle \widetilde{C}_{n}(q) \rangle_{M}=\langle C^{i,i}_{n}(q) \rangle_{M}-\langle C^{i,j}_{n}(q) \rangle_{M}.
\label{Eq:FCDiffaver}
\end{equation}
Such difference spectrum facilitates identification of Fourier components that are caused only by the sample structure.

\section{Experimental}

The coherent x-ray scattering experiment was performed at the nanoprobe endstation GINIX \cite{Kalbfleisch}
installed at the coherence beamline P10 of the PETRA III facility at DESY in Hamburg.
The scattering geometry of the experiment is shown in Fig.~\ref{Fig:ExpGeom}(a).
The incident photon energy was chosen to be $13\;\rm{keV}$ and a 2D detector was positioned
in transmission geometry at $195\;\rm{mm}$ distance from the sample and protected by a beamstop of $34\;\rm{mm}$ in diameter.
The scattering data were recorded on a hybrid-pixel detector Pilatus 1M from Dectris with $981\times1043$ pixels and a pixel size of $172\times 172\;\mu\rm{m}^{2}$.

\begin{figure}[htb]
\centering
\includegraphics[width=8.3cm]{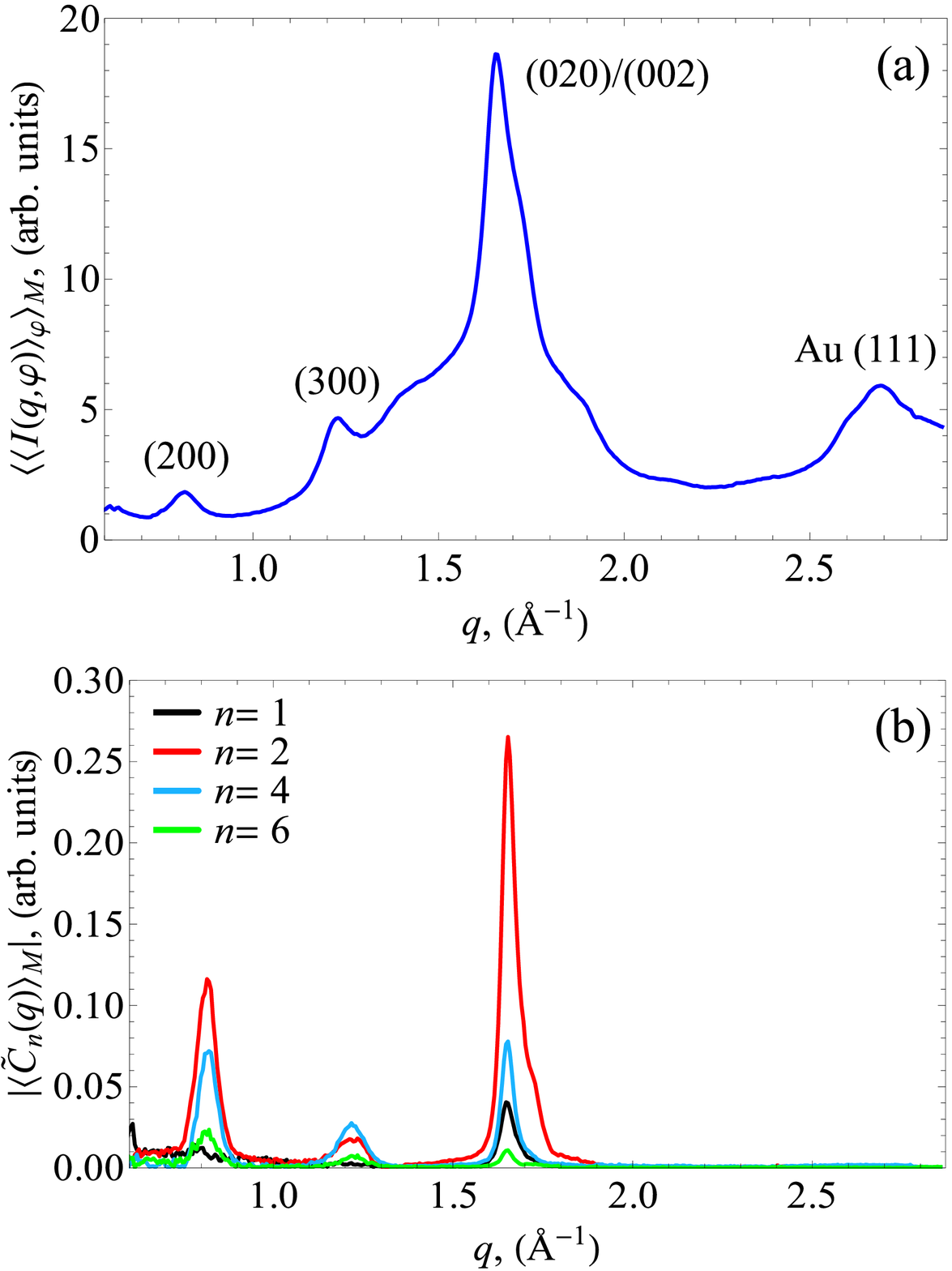}
\caption{(a) Ensemble averaged radial intensity $\langle \langle I(q,\varphi) \rangle_{\varphi}\rangle_{M}$. (b)
Difference Fourier components $\langle \widetilde{C}_{n}(q) \rangle_{M}$ as a function of $q$.
Only the Fourier components of the orders $n=1,2,4$ and 6 are shown.
\label{Fig:SpectraAveraged}}
\end{figure}

The sample was mounted on a goniometer, and a film was aligned with its surface perpendicular to the direction of the incident beam.
The beam with a flux of about $5\cdot10^{10}\;\rm{photons/sec}$ was focused on the sample with Kirkpatrick-Baez (KB) mirrors to a spot of about $200\times300\;\rm{nm}^2$ (FWHM).
The sample was scanned in the $x-y$ plane perpendicular to the incident beam direction.
The sample area of $20\times 40\;\mu\rm{m}^{2}$ was scanned on the $20\times40$ raster grid with a $1\;\mu\rm{m}$ step size in both directions and the total number of $M=800$ positions.
Cryogenic cooling of the film with liquid nitrogen was used during measurements to reduce radiation damage.
The exposure time was chosen to be $0.3\;\rm{sec}$ per image to perform measurements in a nondestructive regime.

The sample was spin-cast from a blend of P3HT (molecular weight $44.9\;\rm{kg/mol}$; PDI 1.22) with gold nanoparticles.
Gold nanoparticles of $2-3\;\rm{nm}$ in size stabilized with fluorene derivatives (AuNPs-SFL) have been prepared and characterized in analogy to recent reports \cite{Fratoddi, Matassa, Quintiliani, Vitaliano, Cametti}.
A 5~mg/ml blend solution (P3HT:AuNPs-SFL = 10:1 by weight) in chloroform was spin cast
on rectangular shaped grids with a 15 nm thick $\rm{Si}_{3}\rm{N}_{4}$ membranes (Dune Sciences, Inc.).

An x-ray dataset consisting of $M=800$ diffraction patterns was measured.
The data were corrected for background scattering and polarization of incident x-rays.
Typical snapshot diffraction patterns corresponding to two different positions on the sample are shown in Figs.~\ref{Fig:ExpGeom}(b)-\ref{Fig:ExpGeom}(d).
It is readily seen that the film is inhomogeneous, i.e. it has different structures at distinct sample positions.

\section{Results and discussion}

We first determine the average structure of the film and then continue with spatially resolved analysis of
nanoscale variations of the film properties.

\subsection{Spatially averaged structural properties}

To determine the average structure of the film we first analyzed the ensemble-averaged intensity.
The radial intensity $\langle \langle I(q,\varphi) \rangle_{\varphi}\rangle_{M}$ averaged
over the full dataset $M$ in the range of $0.6\;\mathring{A}^{-1} \leq q \leq 2.9\;\mathring{A}^{-1}$ is presented in Fig.~\ref{Fig:SpectraAveraged}(a).
Our diffraction data suggest that the average structure of crystalline domains can be described with a monoclinic unit cell
with parameters $a=15.7\;\mathring{A}$ and $b=c=7.6\;\mathring{A}$, similar to the model proposed in \cite{Kayunkid}.
One can clearly see in Fig.~\ref{Fig:SpectraAveraged}(a) the $(200)$ and  $(300)$ peaks at $q=0.82\;\mathring{A}^{-1}$ and $q=1.2\;\mathring{A}^{-1}$, respectively \footnote{The $(100)$ reflection was covered by the beamstop.}.
These peaks are defined by the unit cell parameter $a$ of crystalline P3HT domains [see Fig.~\ref{Fig:P3HT}], and suggest the presence of the face-on
morphology in the sample.
The strongest peak at $q=1.65\;\mathring{A}^{-1}$ may contain scattering contribution from both $c-$ and $b-$planes,
since the $c$ and $b$ lattice parameters have very close values \cite{Grigorian1}.
Due to the fact that crystalline domains may have different orientations in the illuminated sample area, this peak is often considered as $(020)/(002)$ \cite{Joshi}.
Finally, the peak located at $q=2.68\;\mathring{A}^{-1}$ is defined by scattering from the $(111)$ set of atomic planes of gold.
The corresponding scattering ring at higher momentum transfer values is partially visible on the diffraction patterns shown in Figs.~\ref{Fig:ExpGeom}(b)-\ref{Fig:ExpGeom}(d).

To determine the average characteristics of orientational order of P3HT domains,
the Fourier spectra $\langle C^{i,i}_{n}(q) \rangle_{M}$ (see section 1.1 in the ESI$^\dag$) and $\langle C^{i,j}_{n}(q) \rangle_{M}$,
as well as their difference $\langle \widetilde{C}_{n}(q) \rangle_{M}$ (Eq.~\ref{Eq:FCDiffaver})
were calculated and averaged over the full dataset $M$ [see Fig.~\ref{Fig:SpectraAveraged}(b)].
Only Fourier components of the orders $n=1,2,4$ and 6 have large contribution to the full spectrum (the others have negligible values).
The $q$-dependence of $\langle \widetilde{C}_{n}(q) \rangle_{M}$ has three local maxima located at the same $q$ values as the $(200)$, $(300)$ and $(020)/(002)$
peaks on the average intensity curve $\langle \langle I(q,\varphi) \rangle_{\varphi}\rangle_{M}$ [Fig.~\ref{Fig:SpectraAveraged}(a)].
The presence of harmonics higher than $n=2$ (up to $n=6$) indicates enhanced orientational order of P3HT domains.
The nonzero value of the first order Fourier component can be attributed to uncompensated background contribution.

We would like to note, that analysis of diffraction data measured from pristine P3HT films prepared
by the same fabrication protocol does not reveal orientational order in the system.
For pristine P3HT films all Fourier components of the difference spectra $\langle \widetilde{C}_{n}(q) \rangle_{M}$ have negligible values.
This brings us to the conclusion that in P3HT/AuNPs blends orientational order of crystalline domains is induced by Au nanoparticles.
This observation is supported by recent experiments, where improved crystallinity and orientational order were reported for various P3HT blends \cite{Parashchuk, Wu}.

The difference spectrum shown in Fig.~\ref{Fig:SpectraAveraged}(b) is determined by {\it average nanoscale}
structural properties of the sample due to the nano-sized x-ray probe applied in the experiment.
At the same time, the diffraction patterns shown in Figs.~\ref{Fig:ExpGeom}(b) and \ref{Fig:ExpGeom}(c)
clearly indicate structural variations in different parts of the sample.
The major difference between these diffraction patterns is determined by the $(200)$ and $(300)$ peaks present in low $q$-region,
that can be attributed to the face-on orientation of P3HT domains [Fig.~\ref{Fig:ExpGeom}(b),(c)].
Both peaks are absent in Fig.~\ref{Fig:ExpGeom}(d) where only the $(020)/(002)$ scattering ring can be observed, that is characteristic for the edge-on orientation.
Therefore, it is important to characterize the structure of the polymer film at each separate position defined by the x-ray probe.

\subsection{Spatially resolved structural properties}

\begin{figure*}[htb]
\centering
\includegraphics[width=14cm]{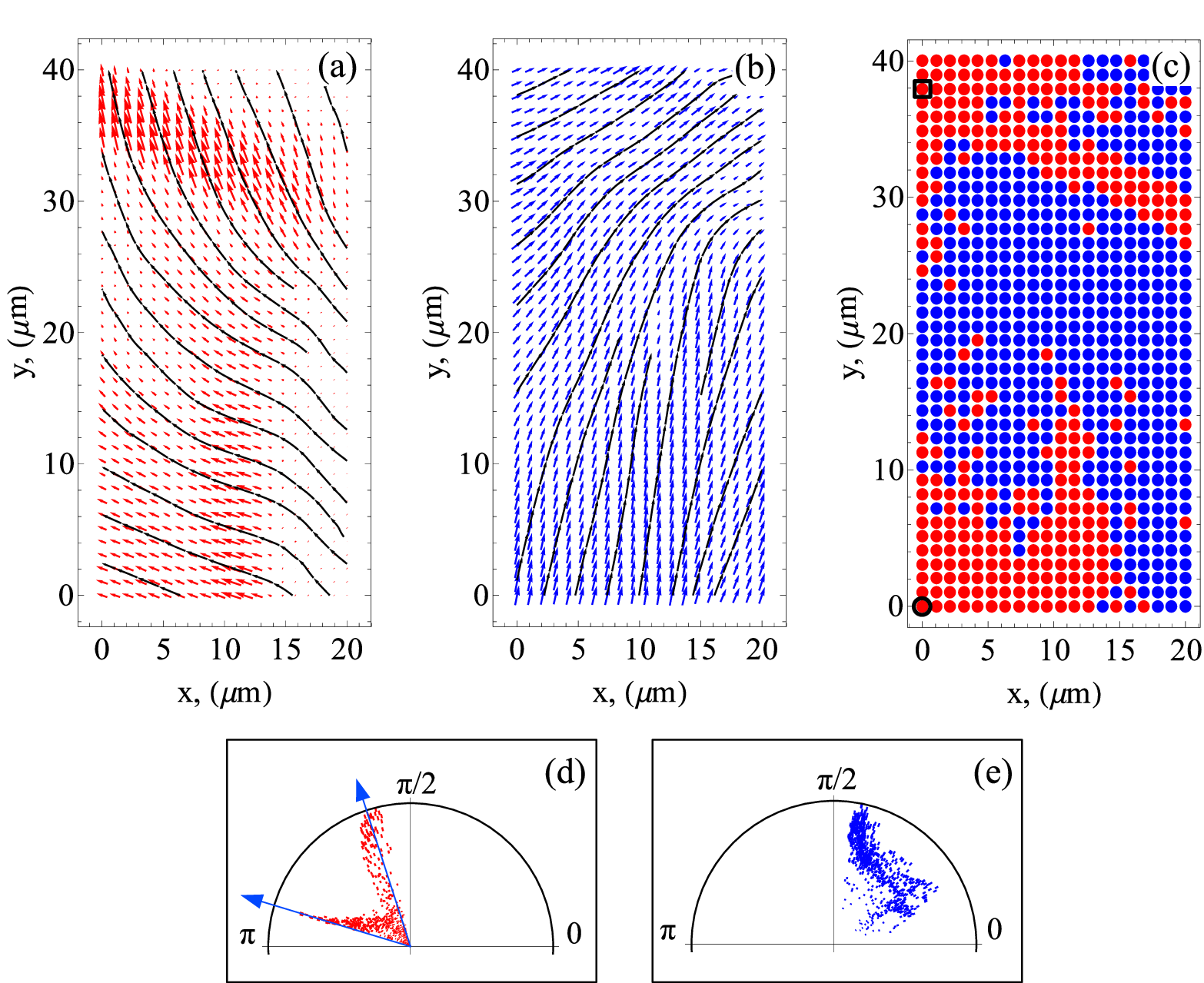}
\caption{Spatial distribution of orientations of crystalline P3HT domains across the film, determined
by the phases $\psi_{2}^{(200)}$ and $\psi_{2}^{(020)/(002)}$ of the Fourier components $I_{2}(q_{200})$ (a) and $I_{2}(q_{020/002})$ (b), respectively.
Each vector in the plotted vector fields corresponds to a certain position on the sample, with the
vector length proportional to the modulus of the respective Fourier component.
Black lines are plotted to guide the eye.
(c) Two sample areas with predominant face-on (red points) and mixed (blue points) orientation of crystalline domains.
Black square and circle in (c) indicate two sample positions chosen for analysis in Fig.~\ref{Fig:SpectraSingleDp}.
Angular distribution of all observed orientations in the vector fields (a) and (b) is shown in polar coordinates in (d) and (e), respectively.
Arrows in (d) specify two major in-plane orientations of the face-on domains in the sample.
\label{Fig:VectorField}}
\end{figure*}

To determine spatially resolved nanoscale information about the structure of the film we analyzed diffraction patterns and corresponding Fourier spectra of intensity $I_{n}(q)$
at each spatial position of the sample.
To analyze local orientational distribution of domains we directly applied Eq.~(\ref{Eq:Cchi}) and determined the phases $\psi_{2}$ of the dominant second order Fourier components $I_{2}(q)$.
This Fourier component has the largest contribution in the spectrum and its phase determines local orientation of P3HT domains in the film.
We determine the distribution of orientations of crystalline P3HT domains across the sample using
two-dimensional (2D) vector fields shown in Figs.~\ref{Fig:VectorField}(a) and \ref{Fig:VectorField}(b).
Each vector in these figures is defined by the amplitude and phase of the
Fourier component $I_{2}(q)$, determined at the positions 
of the $(200)$ [Fig.~\ref{Fig:VectorField}(a)]  and $(020)/(002)$  [Fig.~\ref{Fig:VectorField}(b)] peaks, respectively.
Figs.~\ref{Fig:VectorField}(d) and ~\ref{Fig:VectorField}(e) represent all available orientations
shown in Figs.~\ref{Fig:VectorField}(a) and \ref{Fig:VectorField}(b) in the form of angular diagrams.

The distribution of the magnitudes and orientations of the vectors in Figs.~\ref{Fig:VectorField}(a) and ~\ref{Fig:VectorField}(b) shows that the film is not uniform.
As it follows from Figs.~\ref{Fig:VectorField}(d) and ~\ref{Fig:VectorField}(e), all phases
are distributed in the angular range of $101^{\circ} \leq\psi_{2}^{(200)}\leq 164^{\circ}$ in Fig.~\ref{Fig:VectorField}(a),
and in the range of $12^{\circ} \leq\psi_{2}^{(020)/(002)}\leq 90^{\circ}$ in Fig.~\ref{Fig:VectorField}(b).
A comparison of vector orientations in Figs.~\ref{Fig:VectorField}(a) and ~\ref{Fig:VectorField}(b) allows us to partition the sample area into two major regions [see Fig.~\ref{Fig:VectorField}(c)]
according to the value of the phase difference $\Delta \psi_{2}=\psi_{2}^{(200)}-\psi_{2}^{(020)/(002)}$ calculated at each spatial position.

The first region [red points in Fig.~\ref{Fig:VectorField}(c)] is characterized with a strong correlation between the
vector orientations in Figs.~\ref{Fig:VectorField}(a) and \ref{Fig:VectorField}(b),
and the phase difference in the range of angles $80^{\circ} \leq \Delta \psi_{2}\leq 90^{\circ}$.
This suggests a presence of a certain preferential P3HT morphology in this region,
that preserves angular orientation of the $(200)$ and $(020)/(002)$ peaks with respect to each other at each particular position.
Since our samples were prepared in the same way as in Ref.~\cite{Salammal}, where the face-on oriented domains were mainly observed,
we conclude that the face-on morphology is preferential in this region.
In this case, the main contribution to the peak at $q=1.65\;\mathring{A}^{-1}$ is defined by the $c$-planes [i.e., by the $(002)$ peak].
As one can see in Fig.~\ref{Fig:VectorField}(c), two separated areas of the sample are characterized with dominating face-on morphology
but different orientations, indicated with two arrows in Fig.~\ref{Fig:VectorField}(d).
Quite naturally these regions coincide with the areas in Fig.~\ref{Fig:VectorField}(a) where the magnitudes $|I_{2}(q_{200})|$ are large
since the face-on morphology in the given scattering geometry is primarily associated with a $(200)$ peak.

The second region of the film [blue points in Fig.~\ref{Fig:VectorField}(c)] is characterized with a larger spread of the phase difference, $50^{\circ} \leq \Delta \psi_{2}\leq 80^{\circ}$.
One can see in Figs.~\ref{Fig:VectorField}(a) and ~\ref{Fig:VectorField}(b), that both vector fields are characterized with a smooth change of the magnitudes and orientations of the vectors.
At the same time, distribution of the magnitudes of the vectors in Fig.~\ref{Fig:VectorField}(b) is more uniform than in Fig.~\ref{Fig:VectorField}(a).
It is about $50\%$ of the vectors in Fig.~\ref{Fig:VectorField}(a) have magnitudes smaller than $1/3$ of the largest vector,
while less than $5\%$ of the vectors in Fig.~\ref{Fig:VectorField}(b) satisfy similar condition.
This all suggests that in the second region a mixed orientation of domains is observed, and a continuous transition from the face-on towards the edge-on morphology is achieved by rotation of the face-on domains
around the $c$-axis [see Fig.~\ref{Fig:P3HT}].
It is important to note, that such a conclusion can be drawn only from a simultaneous analysis of both vector fields in Figs.~\ref{Fig:VectorField}(a) and ~\ref{Fig:VectorField}(b),
while separate analysis of each peak does not reflect a real distribution of domain types.

\begin{figure*}[htb]
\centering
\includegraphics[width=15cm]{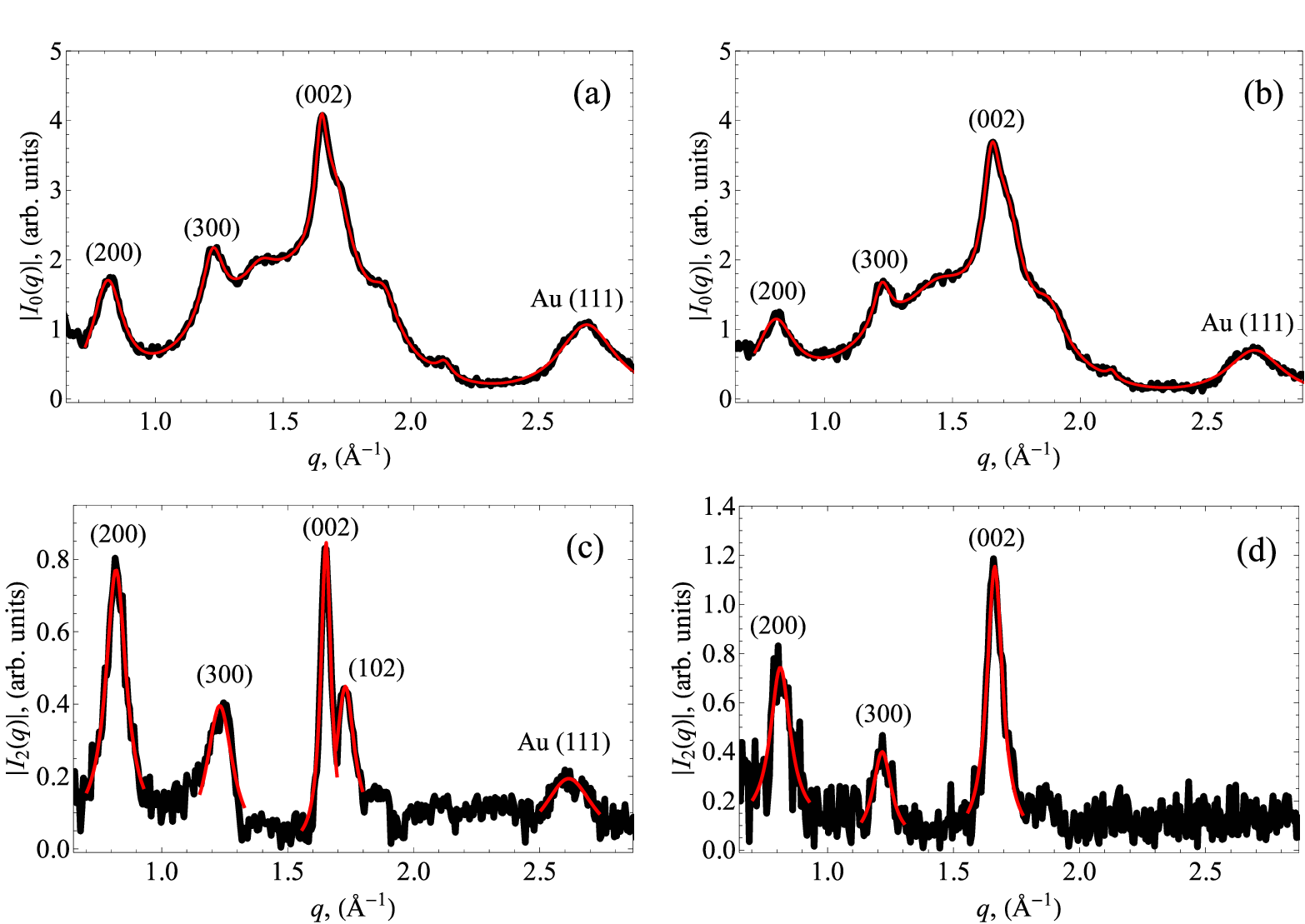}
\caption{Fourier components of intensity of the orders $n=0$ (a),(b) and $n=2$ (c),(d) as a function of $q$,
calculated for two diffraction patterns measured at two sample positions with face-on morphology.
Curves (a) and (c) correspond to the position indicated by a black square in Fig.~\ref{Fig:VectorField}(c), and (b) and (d) correspond to the position marked with a circle in Fig.~\ref{Fig:VectorField}(c).
Black and red curves correspond to the experimental data and results of Lorentzian fits, respectively (see text).
In the case of $n=0$ a linear background contribution has been subtracted.
\label{Fig:SpectraSingleDp}}
\end{figure*}
%

A smooth distribution of the phases and magnitudes of the Fourier components $I_{2}(q)$ across the sample shown in Figs.~\ref{Fig:VectorField}(a) and \ref{Fig:VectorField}(b) indicates a
substantial orientational order in the polymer film and confirms applicability of the spatially resolved XCCA to such a system.
As a step further, we applied the CCF $C^{i,i}(q,\Delta)$ to individual diffraction patterns to access hidden structural features of the film with an increased accuracy (see section 1.2 in the ESI$^\dag$).
In this case the Fourier components of intensity were determined using the relation $|I_{n}(q)|=\sqrt{C^{i,i}_{n}(q)}$ (see Eq.~\ref{Eq:Cqnii2}).
In Fig.~\ref{Fig:SpectraSingleDp} the magnitudes $|I_{n}(q)|$ of the Fourier components of the orders $n=0$ and $n=2$ are presented as a function of $q$
for two diffraction patterns measured at different sample positions with the predominant face-on morphology.
One can see that the $q$-dependence of $|I_{2}(q)|$ provides additional structural details that can not be directly observed in the angular averaged profile 
$|I_{0}(q)|\equiv \langle I(q,\varphi) \rangle_{\varphi}$, that is
usually analyzed in scattering experiments.
For example, the peak indexed as $(102)$ on the $|I_{2}(q)|$ profile can be clearly separated from the $(002)$ peak in Fig.~\ref{Fig:SpectraSingleDp}(c),
and is not visible in Fig.~\ref{Fig:SpectraSingleDp}(d), indicating slightly different orientation of the face-on domains at these positions.
This cannot be directly seen from the $|I_{0}(q)|$ profiles shown in Figs.~\ref{Fig:SpectraSingleDp}(a) and ~\ref{Fig:SpectraSingleDp}(b).

We determined the positions of the centers $q_{0}$ and magnitudes $|I_{n}(q_{0})|$ of different peaks in $|I_{0}(q)|$ and $|I_{2}(q)|$ profiles as a function of the probe position on the sample.
These peaks were fitted with a Lorentzian function (see Fig.~\ref{Fig:SpectraSingleDp}),
$|I_{n}(q)|=b+s\cdot\gamma_{n}\cdot[(q-q_{0})^2+\gamma_{n}^2]^{-1}$, where $b$ is a background correction, $s$ is a scaling coefficient,
 $\gamma_{n}$ is half-width at half-maximum (HWHM) of the peak, and $q_{0}$ is a center of the peak.
Analysis of the $(200)$ peak on the $|I_{0}(q)|$ profile (see section 1.3 in the ESI$^\dag$),
gives the average positional correlation length $\xi=2\pi/\gamma_{0}$ of the order of $10\;\rm{nm}$.
This value is in agreement with a characteristic size of P3HT crystalline domains \cite{Salammal, Kohn}.

\begin{figure}[htb]
\centering
\includegraphics[width=8.3cm]{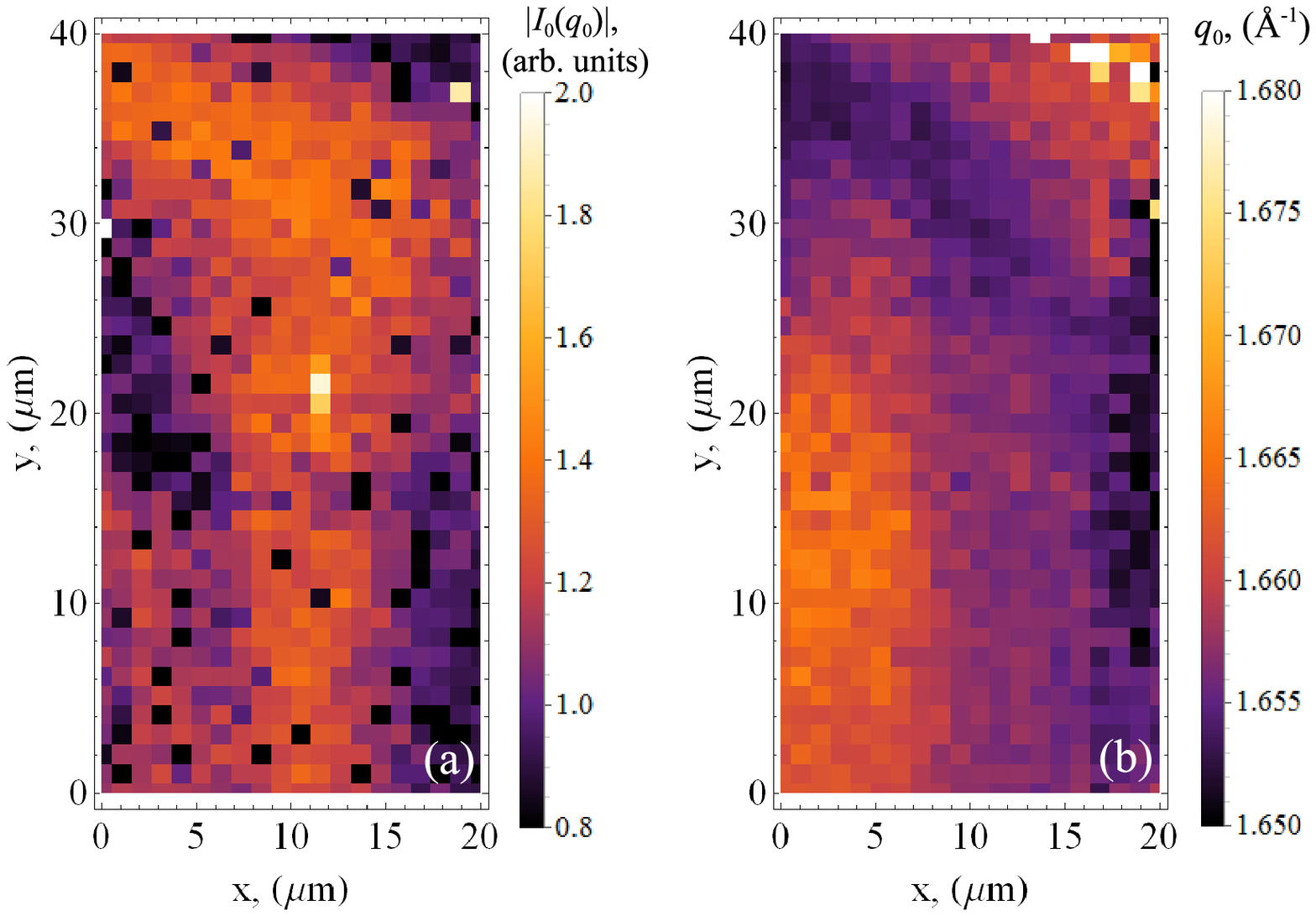}
\caption{Spatially resolved 2D maps of (a) magnitude $|I_{0}(q_{0})|$ of Au~$(111)$ peak on the $|I_{0}(q)|$ profile, (b) position $q_{0}$ of the P3HT $(002)$ peak determined on the $|I_{2}(q)|$ profile.
\label{Fig:Map2D}}
\end{figure}

In Fig.~\ref{Fig:Map2D}(a) a 2D map of the magnitudes $|I_{0}(q_{0})|$ of the Au $(111)$ peak is shown, determined from the fit of the $(111)$ peak on the $|I_{0}(q)|$ profile.
As one can see, gold nanoparticles are not uniformly distributed across the film,
the regions with high gold concentration correspond to large values of Au $(111)$ peak magnitude.
A 2D map of the P3HT $(002)$ peak center position $q_{0}$ determined from the $|I_{2}(q)|$ profile is shown in Fig.~\ref{Fig:Map2D}(b).
One can clearly see a smooth variation of the $(002)$ peak position,
that indicates changes of the unit cell parameter $c$ across the film in the range from $7.54\;\mathring{A}$ to $7.62\;\mathring{A}$.
In the regions with higher gold concentration the $(002)$ peak slightly shifts to lower $q$.
This observation suggests that Au nanoparticles could influence the structure of P3HT crystalline regions, particularly cause
relaxation of the $c$-spacing in the P3HT backbone [see Fig.~\ref{Fig:P3HT}].

\section{Conclusions}

The x-ray scattering experiment with a nanofocused beam in combination with XCCA allowed us to reveal details
of the nanoscale structure of semicrystalline films cast from P3HT/AuNPs blend.
We explored spatially resolved and average properties of the films by 2D mapping various structural parameters.
One of the prominent results obtained in our work is orientational distribution of crystalline domains
that allowed us to distinguish sample regions with predominant face-on and mixed orientation of domains.
Our results suggest that a continuous transition between different morphologies is achieved by rotation of the domains around the $c$-axis defined along the P3HT backbone.

The polymer film is characterized by substantial orientational ordering of crystalline domains,
that is defined by the presence of higher order ($n=2,4$ and $6$) Fourier components of the cross-correlation function.
As a result of our analysis we conclude that orientational order of P3HT domains can be induced by Au nanoparticles.
Spatially resolved 2D maps show inhomogeneities in the spatial distribution of Au nanoparticles, where the regions with a high concentration of AuNPs
coincide with regions where structural relaxation of P3HT matrix is observed.
The average size of crystalline domains was determined to be of the order of $10\;\rm{nm}$.

The obtained results demonstrate that XCCA provides valuable information about the structure of partially ordered materials, complementary
to the conventional SAXS or GIXD analysis.
Spatially resolved analysis of the Fourier spectra of CCFs allows to observe structural features hidden in the averaged SAXS intensity, and
to determine nanoscale variation and interplay between different film parameters. 
It is an irreplaceable tool to observe local structural changes in P3HT conjugated network induced by Au nanoparticles.
As shown here, XCCA has a significant potential to be used as sensitive tool for nanoscale characterization of nanocomposite materials.

\section{Acknowledgements}
We are grateful to S. Allard and U. Scherf from the University of Wuppertal, Germany for synthesis of the P3HT material.
We are thankful to the group of T. Salditt, especially to M. Osterhoff and S. Kalbfleisch,
for providing support of the G{\"o}ttingen instrument for nano-imaging with x-rays (GINIX),
and also A. Zozulya for the help during the experiment. We acknowledge fruitful discussions and support of this project by E. Weckert.
We gratefully acknowledge U. Pietsch for helpful discussions.
Part of this work was supported by BMBF Proposal 05K10CHG ``Coherent Diffraction Imaging and Scattering of Ultrashort Coherent Pulses with Matter''
in the framework of the German-Russian collaboration ``Development and Use of Accelerator-Based Photon Sources''
and the Virtual Institute VH-VI-403 of the Helmholtz Association, the Department of Chemistry of the
Sapienza University of Rome through the Supporting Research Initiative 2013 and BMBF (project Nr 05K3PS4).





\begin{mcitethebibliography}{53}
\providecommand*{\natexlab}[1]{#1}
\providecommand*{\mciteSetBstSublistMode}[1]{}
\providecommand*{\mciteSetBstMaxWidthForm}[2]{}
\providecommand*{\mciteBstWouldAddEndPuncttrue}
  {\def\EndOfBibitem{\unskip.}}
\providecommand*{\mciteBstWouldAddEndPunctfalse}
  {\let\EndOfBibitem\relax}
\providecommand*{\mciteSetBstMidEndSepPunct}[3]{}
\providecommand*{\mciteSetBstSublistLabelBeginEnd}[3]{}
\providecommand*{\EndOfBibitem}{}
\mciteSetBstSublistMode{f}
\mciteSetBstMaxWidthForm{subitem}
{(\emph{\alph{mcitesubitemcount}})}
\mciteSetBstSublistLabelBeginEnd{\mcitemaxwidthsubitemform\space}
{\relax}{\relax}

\bibitem[Kline \emph{et~al.}(2005)Kline, McGehee, Kadnikova, Liu, Fr{\'e}che,
  and Toney]{Kline2}
R.~J. Kline, M.~D. McGehee, E.~N. Kadnikova, J.~Liu, J.~M.~J. Fr{\'e}che and
  M.~F. Toney, \emph{Macromolecules}, 2005, \textbf{38}, 3312\relax
\mciteBstWouldAddEndPuncttrue
\mciteSetBstMidEndSepPunct{\mcitedefaultmidpunct}
{\mcitedefaultendpunct}{\mcitedefaultseppunct}\relax
\EndOfBibitem
\bibitem[Majewski \emph{et~al.}(2006)Majewski, Kingsley, Balocco, and
  Song]{Majewski}
L.~A. Majewski, J.~W. Kingsley, C.~Balocco and A.~M. Song, \emph{Appl. Phys.
  Lett.}, 2006, \textbf{88}, 222108\relax
\mciteBstWouldAddEndPuncttrue
\mciteSetBstMidEndSepPunct{\mcitedefaultmidpunct}
{\mcitedefaultendpunct}{\mcitedefaultseppunct}\relax
\EndOfBibitem
\bibitem[Coakley and McGehee(2004)]{Coakley}
K.~Coakley and M.~McGehee, \emph{Appl. Phys. Lett.}, 2004, \textbf{16},
  4533\relax
\mciteBstWouldAddEndPuncttrue
\mciteSetBstMidEndSepPunct{\mcitedefaultmidpunct}
{\mcitedefaultendpunct}{\mcitedefaultseppunct}\relax
\EndOfBibitem
\bibitem[Sirringhaus \emph{et~al.}(1999)Sirringhaus, Brown, Friend, Nielsen,
  Bechgaard, Langeveld-Voss, Spiering, Janssen, Meijer, Herwig, and
  de~Leeuw]{Sirring}
H.~Sirringhaus, P.~J. Brown, R.~H. Friend, M.~M. Nielsen, K.~Bechgaard,
  B.~M.~W. Langeveld-Voss, A.~J.~H. Spiering, R.~A.~J. Janssen, E.~W. Meijer,
  P.~Herwig and D.~M. de~Leeuw, \emph{Nature}, 1999, \textbf{401}, 685\relax
\mciteBstWouldAddEndPuncttrue
\mciteSetBstMidEndSepPunct{\mcitedefaultmidpunct}
{\mcitedefaultendpunct}{\mcitedefaultseppunct}\relax
\EndOfBibitem
\bibitem[Salleo \emph{et~al.}(2010)Salleo, Kline, Delongchamp, and
  Chabinyc]{Salleo}
A.~Salleo, R.~J. Kline, D.~M. Delongchamp and M.~L. Chabinyc, \emph{Adv.
  Mater.}, 2010, \textbf{22}, 3812\relax
\mciteBstWouldAddEndPuncttrue
\mciteSetBstMidEndSepPunct{\mcitedefaultmidpunct}
{\mcitedefaultendpunct}{\mcitedefaultseppunct}\relax
\EndOfBibitem
\bibitem[Dang \emph{et~al.}(2011)Dang, Hirsch, and Wantz]{Dang}
M.~T. Dang, L.~Hirsch and G.~Wantz, \emph{Adv. Mater.}, 2011, \textbf{23},
  3597\relax
\mciteBstWouldAddEndPuncttrue
\mciteSetBstMidEndSepPunct{\mcitedefaultmidpunct}
{\mcitedefaultendpunct}{\mcitedefaultseppunct}\relax
\EndOfBibitem
\bibitem[Krebs \emph{et~al.}(2013)Krebs, Espinosa, H{\"o}sel, Sondergaard, and
  Jorgensen]{Krebs}
F.~C. Krebs, N.~Espinosa, M.~H{\"o}sel, R.~R. Sondergaard and M.~Jorgensen,
  \emph{Adv. Mater.}, 2013, \textbf{26}, 29\relax
\mciteBstWouldAddEndPuncttrue
\mciteSetBstMidEndSepPunct{\mcitedefaultmidpunct}
{\mcitedefaultendpunct}{\mcitedefaultseppunct}\relax
\EndOfBibitem
\bibitem[Y.~Kim \emph{et~al.}(2006)Y.~Kim, Cook, Tuladhar, Choulis, Durrant,
  Bradley, Giles, McCulloch, Ha, and Ree]{Kim1}
Y.~Y.~Kim, S.~Cook, S.~M. Tuladhar, S.~A. Choulis, J.~N. J.~R. Durrant,
  D.~D.~C. Bradley, M.~Giles, I.~McCulloch, C.-S. Ha and M.~A. Ree,
  \emph{Macromolecules}, 2006, \textbf{39}, 5843\relax
\mciteBstWouldAddEndPuncttrue
\mciteSetBstMidEndSepPunct{\mcitedefaultmidpunct}
{\mcitedefaultendpunct}{\mcitedefaultseppunct}\relax
\EndOfBibitem
\bibitem[Nagarjuna and Venkataraman(2012)]{Nagarjuna}
G.~Nagarjuna and D.~Venkataraman, \emph{J. Polym. Sci.}, 2012, \textbf{50},
  1045\relax
\mciteBstWouldAddEndPuncttrue
\mciteSetBstMidEndSepPunct{\mcitedefaultmidpunct}
{\mcitedefaultendpunct}{\mcitedefaultseppunct}\relax
\EndOfBibitem
\bibitem[Kohn \emph{et~al.}(2012)Kohn, H{\"u}ttner, Komber, Senkovskyy,
  Tkachov, Kiriy, Friend, Steiner, Huck, Sommer, and Sommer]{Kohn1}
P.~Kohn, S.~H{\"u}ttner, H.~Komber, V.~Senkovskyy, R.~Tkachov, A.~Kiriy, R.~H.
  Friend, U.~Steiner, W.~T.~S. Huck, J.-U. Sommer and M.~Sommer, \emph{J. Am.
  Chem. Soc.}, 2012, \textbf{134}, 4790\relax
\mciteBstWouldAddEndPuncttrue
\mciteSetBstMidEndSepPunct{\mcitedefaultmidpunct}
{\mcitedefaultendpunct}{\mcitedefaultseppunct}\relax
\EndOfBibitem
\bibitem[Brinkmann and Rannou(2007)]{Brinkmann}
M.~Brinkmann and P.~Rannou, \emph{Adv. Funct. Mater.}, 2007, \textbf{17},
  101\relax
\mciteBstWouldAddEndPuncttrue
\mciteSetBstMidEndSepPunct{\mcitedefaultmidpunct}
{\mcitedefaultendpunct}{\mcitedefaultseppunct}\relax
\EndOfBibitem
\bibitem[Kline \emph{et~al.}(2003)Kline, McGehee, Kadnikova, Liu, and
  Fr{\'e}chet]{Kline1}
R.~J. Kline, M.~D. McGehee, E.~N. Kadnikova, J.~Liu and J.~M.~J. Fr{\'e}chet,
  \emph{Adv. Mater.}, 2003, \textbf{15}, 1519\relax
\mciteBstWouldAddEndPuncttrue
\mciteSetBstMidEndSepPunct{\mcitedefaultmidpunct}
{\mcitedefaultendpunct}{\mcitedefaultseppunct}\relax
\EndOfBibitem
\bibitem[A.~Zen \emph{et~al.}(2004)A.~Zen, Pflaum, Hirschmann, Zhuang, Jaiser,
  Asawapirom, Rabe, Scherf, and Neher]{Zen}
A.~A.~Zen, J.~Pflaum, S.~Hirschmann, W.~Zhuang, F.~Jaiser, U.~Asawapirom, J.~P.
  Rabe, U.~Scherf and D.~Neher, \emph{Adv. Funct. Mater.}, 2004, \textbf{14},
  757\relax
\mciteBstWouldAddEndPuncttrue
\mciteSetBstMidEndSepPunct{\mcitedefaultmidpunct}
{\mcitedefaultendpunct}{\mcitedefaultseppunct}\relax
\EndOfBibitem
\bibitem[Ali \emph{et~al.}(2013)Ali, Pietsch, and Grigorian]{Ali}
K.~Ali, U.~Pietsch and S.~Grigorian, \emph{J. Appl. Cryst.}, 2013, \textbf{46},
  908\relax
\mciteBstWouldAddEndPuncttrue
\mciteSetBstMidEndSepPunct{\mcitedefaultmidpunct}
{\mcitedefaultendpunct}{\mcitedefaultseppunct}\relax
\EndOfBibitem
\bibitem[Tanigaki \emph{et~al.}(2014)Tanigaki, Ikeo, Mizokuro, Heck, and
  Aota]{Tanigaki}
N.~Tanigaki, Y.~Ikeo, T.~Mizokuro, C.~Heck and H.~Aota, \emph{Jap. J. Appl.
  Phys.}, 2014, \textbf{53}, 01AB05\relax
\mciteBstWouldAddEndPuncttrue
\mciteSetBstMidEndSepPunct{\mcitedefaultmidpunct}
{\mcitedefaultendpunct}{\mcitedefaultseppunct}\relax
\EndOfBibitem
\bibitem[Joshi \emph{et~al.}(2008)Joshi, Grigorian, Pietsch, Pingel, Zen,
  Neher, and Scherf]{Joshi}
S.~Joshi, S.~Grigorian, U.~Pietsch, P.~Pingel, A.~Zen, D.~Neher and U.~Scherf,
  \emph{Macromolecules}, 2008, \textbf{41}, 6800\relax
\mciteBstWouldAddEndPuncttrue
\mciteSetBstMidEndSepPunct{\mcitedefaultmidpunct}
{\mcitedefaultendpunct}{\mcitedefaultseppunct}\relax
\EndOfBibitem
\bibitem[Prosa \emph{et~al.}(1992)Prosa, Winokur, Moulton, Smith, and
  Heeger]{Prosa}
T.~J. Prosa, M.~J. Winokur, J.~Moulton, P.~Smith and A.~J. Heeger,
  \emph{Macromolecules}, 1992, \textbf{25}, 4364\relax
\mciteBstWouldAddEndPuncttrue
\mciteSetBstMidEndSepPunct{\mcitedefaultmidpunct}
{\mcitedefaultendpunct}{\mcitedefaultseppunct}\relax
\EndOfBibitem
\bibitem[Rahimi \emph{et~al.}(2012)Rahimi, Botiz, Stingelin, Kayunkid, Sommer,
  Koch, Nguyen, Coulembier, Dubois, Brinkmann, and Reiter]{Rahimi}
K.~Rahimi, I.~Botiz, N.~Stingelin, N.~Kayunkid, M.~Sommer, F.~P. Koch,
  H.~Nguyen, O.~Coulembier, P.~Dubois, M.~Brinkmann and G.~Reiter, \emph{Angew.
  Chem. Int. Ed. Engl.}, 2012, \textbf{51}, 11131\relax
\mciteBstWouldAddEndPuncttrue
\mciteSetBstMidEndSepPunct{\mcitedefaultmidpunct}
{\mcitedefaultendpunct}{\mcitedefaultseppunct}\relax
\EndOfBibitem
\bibitem[Kim \emph{et~al.}(2006)Kim, Jang, Park, and Cho]{Kim}
D.~H. Kim, Y.~Jang, Y.~D. Park and K.~Cho, \emph{Macromolecules}, 2006,
  \textbf{39}, 5843\relax
\mciteBstWouldAddEndPuncttrue
\mciteSetBstMidEndSepPunct{\mcitedefaultmidpunct}
{\mcitedefaultendpunct}{\mcitedefaultseppunct}\relax
\EndOfBibitem
\bibitem[Salammal \emph{et~al.}(2012)Salammal, Mikayelyan, Grigorian, Pietsch,
  Koenen, Scherf, Kayunkid, and Brinkmann]{Salammal}
T.~S. Salammal, E.~Mikayelyan, S.~Grigorian, U.~Pietsch, N.~Koenen, U.~Scherf,
  N.~Kayunkid and M.~Brinkmann, \emph{Macromolecules}, 2012, \textbf{45},
  5575\relax
\mciteBstWouldAddEndPuncttrue
\mciteSetBstMidEndSepPunct{\mcitedefaultmidpunct}
{\mcitedefaultendpunct}{\mcitedefaultseppunct}\relax
\EndOfBibitem
\bibitem[Shabi \emph{et~al.}(2012)Shabi, Grigorian, Brinkmann, Pietsch, Koenen,
  Kayunkid, and Scherf]{Shabi}
T.~S. Shabi, S.~Grigorian, M.~Brinkmann, U.~Pietsch, N.~Koenen, N.~Kayunkid and
  U.~Scherf, \emph{J. Appl. Polym. Sci.}, 2012,  2335\relax
\mciteBstWouldAddEndPuncttrue
\mciteSetBstMidEndSepPunct{\mcitedefaultmidpunct}
{\mcitedefaultendpunct}{\mcitedefaultseppunct}\relax
\EndOfBibitem
\bibitem[Brinkmann and Wittmann(2006)]{Brinkmann1}
M.~Brinkmann and J.~C. Wittmann, \emph{Adv. Mater.}, 2006, \textbf{18},
  860\relax
\mciteBstWouldAddEndPuncttrue
\mciteSetBstMidEndSepPunct{\mcitedefaultmidpunct}
{\mcitedefaultendpunct}{\mcitedefaultseppunct}\relax
\EndOfBibitem
\bibitem[Hoppe and Sariciftci(2004)]{Hoppe}
H.~Hoppe and N.~Sariciftci, \emph{J. Mater. Res.}, 2004, \textbf{19},
  1924\relax
\mciteBstWouldAddEndPuncttrue
\mciteSetBstMidEndSepPunct{\mcitedefaultmidpunct}
{\mcitedefaultendpunct}{\mcitedefaultseppunct}\relax
\EndOfBibitem
\bibitem[Zhang \emph{et~al.}(2006)Zhang, Li, Iovu, Jeffries-EL, Sauve, Cooper,
  Jia, Tristram-Nagle, Smilgies, Lambeth, McCullough, and Kowalewski]{Zhang}
R.~Zhang, B.~Li, C.~M. Iovu, M.~Jeffries-EL, G.~Sauve, J.~Cooper, S.~Jia,
  S.~Tristram-Nagle, D.~M. Smilgies, D.~N. Lambeth, R.~D. McCullough and
  T.~Kowalewski, \emph{J. Am. Chem. Soc.}, 2006, \textbf{128}, 3480\relax
\mciteBstWouldAddEndPuncttrue
\mciteSetBstMidEndSepPunct{\mcitedefaultmidpunct}
{\mcitedefaultendpunct}{\mcitedefaultseppunct}\relax
\EndOfBibitem
\bibitem[Grigorian \emph{et~al.}(2010)Grigorian, Joshi, and
  Pietsch]{Grigorian1}
S.~Grigorian, S.~Joshi and U.~Pietsch, \emph{IOP Conf. Series: Mat. Sci. and
  Engineering}, 2010, \textbf{14}, 012007\relax
\mciteBstWouldAddEndPuncttrue
\mciteSetBstMidEndSepPunct{\mcitedefaultmidpunct}
{\mcitedefaultendpunct}{\mcitedefaultseppunct}\relax
\EndOfBibitem
\bibitem[Kohn \emph{et~al.}(2013)Kohn, Rong, Scherer, Sepe, Sommer,
  M{\"u}ller-Buschbaum, Friend, Steiner, and H{\"u}ttner]{Kohn}
P.~Kohn, Z.~Rong, K.~H. Scherer, A.~Sepe, M.~Sommer, P.~M{\"u}ller-Buschbaum,
  R.~H. Friend, U.~Steiner and S.~H{\"u}ttner, \emph{Macromolecules}, 2013,
  \textbf{46}, 4002\relax
\mciteBstWouldAddEndPuncttrue
\mciteSetBstMidEndSepPunct{\mcitedefaultmidpunct}
{\mcitedefaultendpunct}{\mcitedefaultseppunct}\relax
\EndOfBibitem
\bibitem[Kayunkid \emph{et~al.}(2010)Kayunkid, Uttiya, and Brinkmann]{Kayunkid}
N.~Kayunkid, S.~Uttiya and M.~Brinkmann, \emph{Macromolecules}, 2010,
  \textbf{43}, 4961\relax
\mciteBstWouldAddEndPuncttrue
\mciteSetBstMidEndSepPunct{\mcitedefaultmidpunct}
{\mcitedefaultendpunct}{\mcitedefaultseppunct}\relax
\EndOfBibitem
\bibitem[Newbloom \emph{et~al.}(2012)Newbloom, Weigandt, and Pozzo]{Newbloom}
G.~M. Newbloom, K.~M. Weigandt and D.~C. Pozzo, \emph{Macromolecules}, 2012,
  \textbf{45}, 3452\relax
\mciteBstWouldAddEndPuncttrue
\mciteSetBstMidEndSepPunct{\mcitedefaultmidpunct}
{\mcitedefaultendpunct}{\mcitedefaultseppunct}\relax
\EndOfBibitem
\bibitem[Joshi \emph{et~al.}(2009)Joshi, Pindel, Grigorian, Panzner, Pietsch,
  Neher, Forster, and Scherf]{Joshi1}
S.~Joshi, P.~Pindel, S.~Grigorian, T.~Panzner, U.~Pietsch, D.~Neher, M.~Forster
  and U.~Scherf, \emph{Macromolecules}, 2009, \textbf{42}, 4651\relax
\mciteBstWouldAddEndPuncttrue
\mciteSetBstMidEndSepPunct{\mcitedefaultmidpunct}
{\mcitedefaultendpunct}{\mcitedefaultseppunct}\relax
\EndOfBibitem
\bibitem[Moghaddam \emph{et~al.}(2013)Moghaddam, Huettner, Vaynzof, Ducati,
  Divitini, Lohwasser, Musselman, Sepe, Scherer, Thelakkat, Steiner, and
  Friend]{Moghaddam}
R.~S. Moghaddam, S.~Huettner, Y.~Vaynzof, C.~Ducati, G.~Divitini, R.~H.
  Lohwasser, K.~P. Musselman, A.~Sepe, M.~R.~J. Scherer, M.~Thelakkat,
  U.~Steiner and R.~H. Friend, \emph{Nano Lett.}, 2013, \textbf{13}, 4499\relax
\mciteBstWouldAddEndPuncttrue
\mciteSetBstMidEndSepPunct{\mcitedefaultmidpunct}
{\mcitedefaultendpunct}{\mcitedefaultseppunct}\relax
\EndOfBibitem
\bibitem[Parashchuk \emph{et~al.}(2013)Parashchuk, Grigorian, Levin, Bruevich,
  Bukunova, Golovnin, Dittrich, Dembo, Volkov, and Paraschuk]{Parashchuk}
O.~Parashchuk, S.~Grigorian, E.~Levin, V.~Bruevich, K.~Bukunova, I.~Golovnin,
  T.~Dittrich, K.~Dembo, V.~Volkov and D.~Paraschuk, \emph{J. Phys. Chem.
  Lett}, 2013, \textbf{4}, 1298\relax
\mciteBstWouldAddEndPuncttrue
\mciteSetBstMidEndSepPunct{\mcitedefaultmidpunct}
{\mcitedefaultendpunct}{\mcitedefaultseppunct}\relax
\EndOfBibitem
\bibitem[Quintiliani \emph{et~al.}(2014)Quintiliani, Bassetti, Pasquini,
  Battocchio, Rossi, Mura, Matassa, Fontana, Russo, and Fratoddi]{Quintiliani}
M.~Quintiliani, M.~Bassetti, C.~Pasquini, C.~Battocchio, M.~Rossi, F.~Mura,
  R.~Matassa, L.~Fontana, M.~V. Russo and I.~Fratoddi, \emph{J. Mater. Chem.
  C}, 2014, \textbf{2}, 2517\relax
\mciteBstWouldAddEndPuncttrue
\mciteSetBstMidEndSepPunct{\mcitedefaultmidpunct}
{\mcitedefaultendpunct}{\mcitedefaultseppunct}\relax
\EndOfBibitem
\bibitem[Battocchio \emph{et~al.}(2014)Battocchio, Porcaro, Mukherjee, Magnano,
  Nappini, Fratoddi, Quintiliani, Russo, and Polzonetti]{Battocchio1}
C.~Battocchio, F.~Porcaro, S.~Mukherjee, E.~Magnano, S.~Nappini, I.~Fratoddi,
  M.~Quintiliani, M.~V. Russo and G.~Polzonetti, \emph{J. Phys. Chem. C}, 2014,
  \textbf{118}, 8159\relax
\mciteBstWouldAddEndPuncttrue
\mciteSetBstMidEndSepPunct{\mcitedefaultmidpunct}
{\mcitedefaultendpunct}{\mcitedefaultseppunct}\relax
\EndOfBibitem
\bibitem[Battocchio \emph{et~al.}(2014)Battocchio, Fratoddi, Fontana, Bodo,
  Porcaro, Meneghini, Pis, Nappini, Mobilio, Russo, and
  Polzonetti]{Battocchio2}
C.~Battocchio, I.~Fratoddi, L.~Fontana, E.~Bodo, F.~Porcaro, C.~Meneghini,
  I.~Pis, S.~Nappini, S.~Mobilio, M.~V. Russo and G.~Polzonetti, \emph{Phys.
  Chem. Chem. Phys.}, 2014, \textbf{16}, 11719\relax
\mciteBstWouldAddEndPuncttrue
\mciteSetBstMidEndSepPunct{\mcitedefaultmidpunct}
{\mcitedefaultendpunct}{\mcitedefaultseppunct}\relax
\EndOfBibitem
\bibitem[Venditti \emph{et~al.}(2013)Venditti, Fratoddi, Russo, and
  Bearzotti]{Venditti1}
I.~Venditti, I.~Fratoddi, M.~V. Russo and A.~Bearzotti, \emph{Nanotechnology},
  2013, \textbf{24}, 155503\relax
\mciteBstWouldAddEndPuncttrue
\mciteSetBstMidEndSepPunct{\mcitedefaultmidpunct}
{\mcitedefaultendpunct}{\mcitedefaultseppunct}\relax
\EndOfBibitem
\bibitem[Battocchio \emph{et~al.}(2012)Battocchio, Meneghini, Fratoddi,
  Venditti, Russo, Aquilanti, Maurizio, Bondino, Matassa, Rossi, Mobilio, and
  Polzonetti]{Battocchio3}
C.~Battocchio, C.~Meneghini, I.~Fratoddi, I.~Venditti, M.~V. Russo,
  G.~Aquilanti, C.~Maurizio, F.~Bondino, R.~Matassa, M.~Rossi, S.~Mobilio and
  G.~Polzonetti, \emph{J. Phys. Chem. C}, 2012, \textbf{116}, 19571\relax
\mciteBstWouldAddEndPuncttrue
\mciteSetBstMidEndSepPunct{\mcitedefaultmidpunct}
{\mcitedefaultendpunct}{\mcitedefaultseppunct}\relax
\EndOfBibitem
\bibitem[Ghosh and Pal(2007)]{Ghosh}
S.~K. Ghosh and T.~Pal, \emph{Chem. Rev.}, 2007, \textbf{107}, 4797\relax
\mciteBstWouldAddEndPuncttrue
\mciteSetBstMidEndSepPunct{\mcitedefaultmidpunct}
{\mcitedefaultendpunct}{\mcitedefaultseppunct}\relax
\EndOfBibitem
\bibitem[Clarke and Durrant(2010)]{Clarke}
T.~M. Clarke and J.~R. Durrant, \emph{Chem. Rev.}, 2010, \textbf{110},
  6736\relax
\mciteBstWouldAddEndPuncttrue
\mciteSetBstMidEndSepPunct{\mcitedefaultmidpunct}
{\mcitedefaultendpunct}{\mcitedefaultseppunct}\relax
\EndOfBibitem
\bibitem[Wochner \emph{et~al.}(2009)Wochner, Gutt, Autenrieth, Demmer, Bugaev,
  Diaz-Ortiz, Duri, Zontone, Gr{\"u}bel, and Dosch]{Wochner1}
P.~Wochner, C.~Gutt, T.~Autenrieth, T.~Demmer, V.~Bugaev, A.~Diaz-Ortiz,
  A.~Duri, F.~Zontone, G.~Gr{\"u}bel and H.~Dosch, \emph{Proc. Nat. Acad.
  Sci.}, 2009, \textbf{106}, 11511\relax
\mciteBstWouldAddEndPuncttrue
\mciteSetBstMidEndSepPunct{\mcitedefaultmidpunct}
{\mcitedefaultendpunct}{\mcitedefaultseppunct}\relax
\EndOfBibitem
\bibitem[Altarelli \emph{et~al.}(2010)Altarelli, Kurta, and
  Vartanyants]{Altarelli}
M.~Altarelli, R.~P. Kurta and I.~A. Vartanyants, \emph{Phys. Rev. B}, 2010,
  \textbf{82}, 104207; Erratum: 2012, 86, 179904(E)\relax
\mciteBstWouldAddEndPuncttrue
\mciteSetBstMidEndSepPunct{\mcitedefaultmidpunct}
{\mcitedefaultendpunct}{\mcitedefaultseppunct}\relax
\EndOfBibitem
\bibitem[Kurta \emph{et~al.}(2012)Kurta, Altarelli, Weckert, and
  Vartanyants]{Kurta1}
R.~P. Kurta, M.~Altarelli, E.~Weckert and I.~A. Vartanyants, \emph{Phys. Rev.
  B}, 2012, \textbf{85}, 184204\relax
\mciteBstWouldAddEndPuncttrue
\mciteSetBstMidEndSepPunct{\mcitedefaultmidpunct}
{\mcitedefaultendpunct}{\mcitedefaultseppunct}\relax
\EndOfBibitem
\bibitem[Kurta \emph{et~al.}(2013)Kurta, Dronyak, Altarelli, Weckert, and
  Vartanyants]{Kurta2}
R.~P. Kurta, R.~Dronyak, M.~Altarelli, E.~Weckert and I.~A. Vartanyants,
  \emph{New J. Phys.}, 2013, \textbf{15}, 013059\relax
\mciteBstWouldAddEndPuncttrue
\mciteSetBstMidEndSepPunct{\mcitedefaultmidpunct}
{\mcitedefaultendpunct}{\mcitedefaultseppunct}\relax
\EndOfBibitem
\bibitem[Kurta \emph{et~al.}(2013)Kurta, Altarelli, and Vartanyants]{Kurta3}
R.~P. Kurta, M.~Altarelli and I.~A. Vartanyants, \emph{Adv. Cond. Matt. Phys.},
  2013, \textbf{2013}, 959835\relax
\mciteBstWouldAddEndPuncttrue
\mciteSetBstMidEndSepPunct{\mcitedefaultmidpunct}
{\mcitedefaultendpunct}{\mcitedefaultseppunct}\relax
\EndOfBibitem
\bibitem[Kurta \emph{et~al.}(2013)Kurta, Ostrovskii, Singer, Gorobtsov,
  Shabalin, Dzhigaev, Yefanov, Zozulya, Sprung, and Vartanyants]{Kurta4}
R.~P. Kurta, B.~I. Ostrovskii, A.~Singer, O.~Y. Gorobtsov, A.~Shabalin,
  D.~Dzhigaev, O.~M. Yefanov, A.~V. Zozulya, M.~Sprung and I.~A. Vartanyants,
  \emph{Phys. Rev. E}, 2013, \textbf{88}, 044501\relax
\mciteBstWouldAddEndPuncttrue
\mciteSetBstMidEndSepPunct{\mcitedefaultmidpunct}
{\mcitedefaultendpunct}{\mcitedefaultseppunct}\relax
\EndOfBibitem
\bibitem[Schroer \emph{et~al.}(2014)Schroer, Gutt, and Gr{\"u}bel]{Grubel}
M.~A. Schroer, C.~Gutt and G.~Gr{\"u}bel, \emph{Phys. Rev. E}, 2014,
  \textbf{90}, 012309\relax
\mciteBstWouldAddEndPuncttrue
\mciteSetBstMidEndSepPunct{\mcitedefaultmidpunct}
{\mcitedefaultendpunct}{\mcitedefaultseppunct}\relax
\EndOfBibitem
\bibitem[Liu \emph{et~al.}(2013)Liu, Neish, Stokol, Buckley, Smillie, de~Jonge,
  Ott, Kramer, and Bourgeois]{Liu}
A.~C.~Y. Liu, M.~J. Neish, G.~Stokol, G.~A. Buckley, L.~A. Smillie, M.~D.
  de~Jonge, R.~T. Ott, M.~J. Kramer and L.~Bourgeois, \emph{Phys. Rev. Lett.},
  2013, \textbf{110}, 205505\relax
\mciteBstWouldAddEndPuncttrue
\mciteSetBstMidEndSepPunct{\mcitedefaultmidpunct}
{\mcitedefaultendpunct}{\mcitedefaultseppunct}\relax
\EndOfBibitem
\bibitem[Kurta \emph{et~al.}(2014)Kurta, Grodd, Mikayelyan, Gorobtsov,
  Fratoddi, Venditti, Sprung, Grigorian, and Vartanyants]{Kurta5}
R.~P. Kurta, L.~Grodd, E.~Mikayelyan, O.~Y. Gorobtsov, I.~Fratoddi,
  I.~Venditti, M.~Sprung, S.~Grigorian and I.~A. Vartanyants, \emph{J.Phys:
  Conf.Series}, 2014, \textbf{499}, 012021\relax
\mciteBstWouldAddEndPuncttrue
\mciteSetBstMidEndSepPunct{\mcitedefaultmidpunct}
{\mcitedefaultendpunct}{\mcitedefaultseppunct}\relax
\EndOfBibitem
\bibitem[Kalbfleisch \emph{et~al.}(2011)Kalbfleisch, Neubauer, Kruger, Bartels,
  Osterhoff, Mai, Giewekemeyer, Hartmann, Sprung, and Salditt]{Kalbfleisch}
S.~Kalbfleisch, H.~Neubauer, S.~P. Kruger, M.~Bartels, M.~Osterhoff, D.~D. Mai,
  K.~Giewekemeyer, B.~Hartmann, M.~Sprung and T.~Salditt, \emph{AIP Conf.
  Proc.}, 2011, \textbf{1365}, 96\relax
\mciteBstWouldAddEndPuncttrue
\mciteSetBstMidEndSepPunct{\mcitedefaultmidpunct}
{\mcitedefaultendpunct}{\mcitedefaultseppunct}\relax
\EndOfBibitem
\bibitem[Fratoddi \emph{et~al.}(2011)Fratoddi, Venditti, Battocchio,
  Polzonetti, Bondino, Malvestuto, Piscopiello, Tapfer, and Russo]{Fratoddi}
I.~Fratoddi, I.~Venditti, C.~Battocchio, G.~Polzonetti, F.~Bondino,
  M.~Malvestuto, E.~Piscopiello, L.~Tapfer and M.~V. Russo, \emph{J. Phys.
  Chem. C}, 2011, \textbf{115}, 15198\relax
\mciteBstWouldAddEndPuncttrue
\mciteSetBstMidEndSepPunct{\mcitedefaultmidpunct}
{\mcitedefaultendpunct}{\mcitedefaultseppunct}\relax
\EndOfBibitem
\bibitem[Matassa \emph{et~al.}(2012)Matassa, Fratoddi, Rossi, Battocchio,
  Caminiti, and Russo]{Matassa}
R.~Matassa, I.~Fratoddi, M.~Rossi, C.~Battocchio, R.~Caminiti and M.~V. Russo,
  \emph{J. Phys. Chem. C}, 2012, \textbf{116}, 15795\relax
\mciteBstWouldAddEndPuncttrue
\mciteSetBstMidEndSepPunct{\mcitedefaultmidpunct}
{\mcitedefaultendpunct}{\mcitedefaultseppunct}\relax
\EndOfBibitem
\bibitem[Vitaliano \emph{et~al.}(2009)Vitaliano, Fratoddi, Venditti, Roviello,
  Battocchio, Polzonetti, and Russo]{Vitaliano}
R.~Vitaliano, I.~Fratoddi, I.~Venditti, G.~Roviello, C.~Battocchio,
  G.~Polzonetti and M.~V. Russo, \emph{J. Phys. Chem. A}, 2009, \textbf{113},
  14730\relax
\mciteBstWouldAddEndPuncttrue
\mciteSetBstMidEndSepPunct{\mcitedefaultmidpunct}
{\mcitedefaultendpunct}{\mcitedefaultseppunct}\relax
\EndOfBibitem
\bibitem[Cametti \emph{et~al.}(2011)Cametti, Fratoddi, Venditti, and
  Russo]{Cametti}
C.~Cametti, I.~Fratoddi, I.~Venditti and M.~V. Russo, \emph{Langmuir}, 2011,
  \textbf{27}, 7084\relax
\mciteBstWouldAddEndPuncttrue
\mciteSetBstMidEndSepPunct{\mcitedefaultmidpunct}
{\mcitedefaultendpunct}{\mcitedefaultseppunct}\relax
\EndOfBibitem
\bibitem[Wu \emph{et~al.}(2011)Wu, Jeng, Su, Wei, Su, Chiu, Chen, Su, Su, and
  Su]{Wu}
W.-R. Wu, U.-S. Jeng, C.-J. Su, K.-H. Wei, M.-S. Su, M.-Y. Chiu, C.-Y. Chen,
  W.-B. Su, C.-H. Su and A.-C. Su, \emph{ASC Nano}, 2011, \textbf{5},
  6233\relax
\mciteBstWouldAddEndPuncttrue
\mciteSetBstMidEndSepPunct{\mcitedefaultmidpunct}
{\mcitedefaultendpunct}{\mcitedefaultseppunct}\relax
\EndOfBibitem
\end{mcitethebibliography}

\providecommand*{\mcitethebibliography}{\thebibliography}
\csname @ifundefined\endcsname{endmcitethebibliography}
{\let\endmcitethebibliography\endthebibliography}{}

\end{document}